\documentclass[preprint,showpacs,preprintnumbers,amsmath,amssymb]{revtex4}
\usepackage{graphicx}
\usepackage{dcolumn}
\usepackage{amssymb}
\usepackage{bm}
\usepackage{epsfig}
\usepackage{psfrag}
\usepackage{ulem}

\def\3dots{\:\raisebox{-0.5ex}{$\stackrel{\textstyle.}{:}$}\:}
\def\beq{\begin{equation}}
\def\eeq{\end{equation}}
\def\bea{\begin{eqnarray}}
\def\eea{\end{eqnarray}}

\begin{document}

\title{Low temperature and high pressure Raman and x-ray studies of pyrochlore Tb$_2$Ti$_2$O$_7$ : phonon anomalies and possible phase transition}

\author{Surajit Saha,$^{1}$ D. V. S Muthu,$^{1}$ Surjeet Singh,$^{2}$ B. Dkhil,$^{3}$ R. Suryanarayanan,$^{2}$ G. Dhalenne,$^{2}$ H. K. Poswal,$^4$ S. Karmakar,$^4$ Surinder M. Sharma,$^4$ A. Revcolevschi,$^{2}$}

\author{A. K. Sood$^{1}$}%
 \email{asood@physics.iisc.ernet.in, Phone: +91-80-22932964}
 
\affiliation{$^{1}$Department of Physics, Indian Institute of Science, Bangalore - 560012, India.}
\affiliation{$^{2}$Laboratoire de Physico-Chimie de l'Etat Solide, ICMMO, CNRS, UMR8648, Bat 414, Universite Paris-Sud, 91405 Orsay, France.}%
\affiliation{$^{3}$Laboratoire Structures, Proprietes et Modelisation des Solides, Ecole Centrale Paris, CNRS-UMR8580, Grande voie des vignes, 92295 Chatenay-Malabry Cedex, France}
\affiliation{$^4$Synchrotron Radiation Section, Bhabha Atomic Research Centre, Mumbai 400085, India}

\date{\today}

\begin{abstract}

We have carried out temperature and pressure-dependent Raman and x-ray measurements on single crystals of Tb$_2$Ti$_2$O$_7$. We attribute the observed anomalous temperature dependence of phonons to phonon-phonon anharmonic interactions. The quasiharmonic and anharmonic contributions to the temperature-dependent changes in phonon frequencies are estimated quantitatively using mode Gr\"{u}neisen parameters derived from pressure-dependent Raman experiments and bulk modulus from high pressure x-ray measurements. Further, our Raman and x-ray data suggest a subtle structural deformation of the pyrochlore lattice at $\sim$ 9 GPa. We discuss possible implications of our results on the spin-liquid behaviour of Tb$_2$Ti$_2$O$_7$.

\end{abstract}


\maketitle

\section{Introduction}

Insulating rare-earth titanate pyrochlores (A$_2$Ti$_2$O$_7$, where A is tri-positive rare-earth ion) are known to show complex magnetic behaviors, arising from the geometrical frustration of exchange interaction between the rare-earth spins located on an infinite network of corner-sharing tetrahedrons \cite{ARMS-24-453}. Theoretically, for antiferromagnetically coupled classical or Heisenberg spins on the pyrochlore lattice the magnetic ground state should be infinitely degenerate \cite{PT-59-24}. However, the ubiquitous presence of residual terms, like next near-neighbor interactions, crystal field and dipolar interactions can remove this macroscopic degeneracy either completely or partially leading often to complex spin structures at low temperatures \cite{JAC-408-444}. The only member of the A$_2$Ti$_2$O$_7$ series where the presence of residual terms has apparently no significant influence on the spin-dynamics is Tb$_2$Ti$_2$O$_7$. In this pyrochlore, the strength of antiferromagnetic exchange is of the order of 20 K, however despite this, the Tb$^{3+}$ spins show no signs of freezing or long-range ordering down to a temperature of at least 70 mK \cite{PRL-82-1012}. It has been shown, however, that this ``collective paramagnetic'' or the so-called ``spin-liquid'' state of Tb$^{3+}$ moments is instable under high-pressure \cite{Nature-420-54}. Using powder neutron diffraction experiments, Mirebeau \textit{et al.} \cite{Nature-420-54} showed that application of iso-static pressure of about 8.6 GPa in Tb$_2$Ti$_2$O$_7$ induces a long-range order of Tb spins coexisting with the spin-liquid. Since no indication of pressure induced structural deformation was observed in this study, the spin-crystallization, under pressure, was believed to have resulted from the break-down of a delicate balance among the residual terms.

Recently, the vibrational properties of some of these pyrochlores have been investigated by several groups \cite{APL-86-181906, CPL-413-248, PRB-74-064109, JPCM-17-5225, JRS-39-537, PRB-77-054408, PRB-77-214310, PRB-78-134420, PRB-78-214102}. These studies not only show that phonons in the titanate pyrochlores are highly anomalous, but also indicate the extreme sensitivity of vibrational spectroscopy towards probing subtle structural and electronic features not observed. In the pyrochlore Dy$_2$Ti$_2$O$_7$, Raman spectroscopy revealed a subtle structural deformation of the pyrochlore lattice upon cooling below T = 100 K \cite{PRB-78-214102}. In the pyrochlore Tb$_2$Ti$_2$O$_7$, new crystal-field (CF) excitations were identified using Raman data at T = 4 K \cite{PRB-78-134420}. In the temperature-dependent studies, signature of highly anomalous phonons (i.e., decrease of phonon frequency upon cooling; also referred to as phonon softening) has been witnessed in the pyrochlores Er$_2$Ti$_2$O$_7$ \cite{JRS-39-537}, Gd$_2$Ti$_2$O$_7$ \cite{JRS-39-537, PRB-77-214310} and Dy$_2$Ti$_2$O$_7$ \cite{JPCM-17-5225, JRS-39-537, PRB-77-214310, PRB-78-214102}. The effect of pressure, at ambient temperature, has also been studied recently for several of these titanate pyrochlores. Sm$_2$Ti$_2$O$_7$ and Gd$_2$Ti$_2$O$_7$ pick up anion disorder above 40 GPa and become amorphous above 51 GPa \cite{APL-86-181906, CPL-413-248}. Gd$_2$Ti$_2$O$_7$ exhibits a structural deformation near 9 GPa \cite{PRB-74-064109}.

In this paper we present Raman and powder x-ray diffraction studies on the pyrochlore Tb$_2$Ti$_2$O$_7$. These studies were carried out in the temperature range between room temperature and 27 K; and pressure varying from ambient pressure to 25 GPa. Our study reveals highly anomalous softening of the phonons upon cooling. To understand this anomalous behavior, we have estimated the quasiharmonic contribution to the temperature-dependent shift of the frequencies of different Raman phonons using the mode Gr\"{u}neisen parameters obtained from high-pressure Raman data; and bulk modulus and thermal expansion coefficient obtained from high-pressure and temperature-dependent powder x-ray diffraction data, respectively. These analyses allow us to extract the changes in the phonon frequencies arising solely due to anharmonic interactions. We also bring out the effect of pressure on phonons manifesting a subtle structural deformation of the lattice near 9 GPa which is corroborated by a change in the bulk modulus by $\sim$ 62\%. This observation may have relevance to the observations of powder neutron scattering study \cite{Nature-420-54} mentioned above. While this paper was being written, we came across a very recent temperature-dependent study \cite{PRB-78-134420} on this system revealing phonon softening behavior and the coupling of phonons with the crystal field transitions. We shall compare below our results on the temperature dependence of phonons with those of this recent study and quantify the quasiharmonic and anharmonic contributions to the change in phonon frequencies.

\section{Experimental Techniques}

\subsection{Crystal growth}

Stoichiometric amounts of Tb$_2$O$_3$ (99.99 $\%$) and TiO$_2$ (99.99 $\%$) were mixed thoroughly and heated at 1200 $^\circ$C for about 15 h. The resulting mixture was well ground and isostatically pressed into rods of about 6 cm long and 5 mm diameter. These rods were sintered at 1400 $^\circ$C in air for about 72 h. This procedure was repeated until the compound Tb$_2$Ti$_2$O$_7$ was formed, as revealed by powder x-ray diffraction analysis, with no traces of any secondary phase. These rods were then subjected to single crystal growth by the floating-zone method in an infrared image furnace under flowing oxygen. X-ray diffraction measurement was carried out on the powder obtained by crushing part of a single crystalline sample and energy dispersive x-ray analysis in a scanning electron microscope indicated a pure pyrochlore Tb$_2$Ti$_2$O$_7$ phase. The L\"{a}ue back-reflection technique was used to orient the crystal along the principal crystallographic directions. 

\subsection{Raman measurements}

Raman spectroscopic measurements on a (111) cut thin single-crystalline slice (0.5 mm thick and 3 mm in diameter, polished down to a roughness of almost 10 $\mu$m) of Tb$_2$Ti$_2$O$_7$ were performed at low temperatures in back-scattering geometry, using the 514.5 nm line of an $Ar^+$ ion laser (Spectra-Physics) with $\sim$ 20 mW of power falling on the sample. Temperature scanning was done using a CTI-Cryogenics Closed Cycle Refrigerator. Temperature was measured and controlled (with a maximum error of 0.5 K) using a calibrated Pt-sensor and a CRYO-CON 32B temperature controller. The scattered light was collected by a lens and was analyzed using a computer controlled SPEX Ramalog spectrometer having two holographic gratings (1800 groves/mm) coupled to a Peltier-cooled photo multiplier tube connected to a digital photon counter. 

High-pressure Raman experiments were carried out at room temperature up to $\sim$ 25 GPa in a Mao-Bell type diamond anvil cell (DAC). A single crystalline Tb$_2$Ti$_2$O$_7$ sample (size $\sim$ 50 $\mu$m) was placed with a ruby chip (size $\sim$ 10 $\mu$m) in a hole of $\sim$ 200 $\mu$m diameter drilled in a preindented stainless-steel gasket with a mixture of 4:1 methanol and ethanol as the pressure-transmitting medium. Pressure was calibrated using the ruby fluorescence technique \cite{Science-176-284}.

\subsection{X-ray diffraction}

High resolution x-ray diffraction measurements were performed between 10-300 K (with temperature accuracy better than 0.5 K) using a highly accurate two-axis diffractometer in a Bragg-Brentano geometry (focalization circle of 50 $\mu$m) using the Cu-K$_{\beta}$ line ($\lambda$=1.39223 \AA) of a 18 kW rotating anode. 

For high-pressure x-ray experiments, single crystalline Tb$_2$Ti$_2$O$_7$ samples were crushed into fine powder which was
loaded along with a few particles of copper, in a hole of $\sim$ 120 $\mu$m diameter drilled in a preindented ($\sim$ 70 $\mu$m thick) tungsten gasket of a Mao-Bell-type diamond-anvil cell (DAC). The pressure-transmitting medium was methanol-ethanol-water (16:3:1) mixture, which remains hydrostatic until a pressure of $\sim$ 15 GPa. Pressure was determined from the known equation of state of copper \cite{PRB-70-094112}. High-pressure angle dispersive x-ray diffraction experiments were carried out up to $\sim$ 25 GPa on Tb$_2$Ti$_2$O$_7$ at the 5.2$R$ (XRD1) beamline of the Elettra Synchrotron source (Italy) with monochromatized x-rays ($\lambda $= 0.69012 \AA). The diffraction patterns were recorded using a MAR345 imaging plate detector kept at a distance of $\sim$ 20 cm from the sample. Two-dimensional (2D) imaging plate records were transformed into one-dimensional (1D) diffraction profiles by radial integration of the diffraction rings using the FIT2D software \cite{HPR-14-235}.

\section{Results}

\subsection{Raman spectrum of Tb$_2$Ti$_2$O$_7$}

\begin{figure}
\begin{center}
\leavevmode
\includegraphics[width=\textwidth]{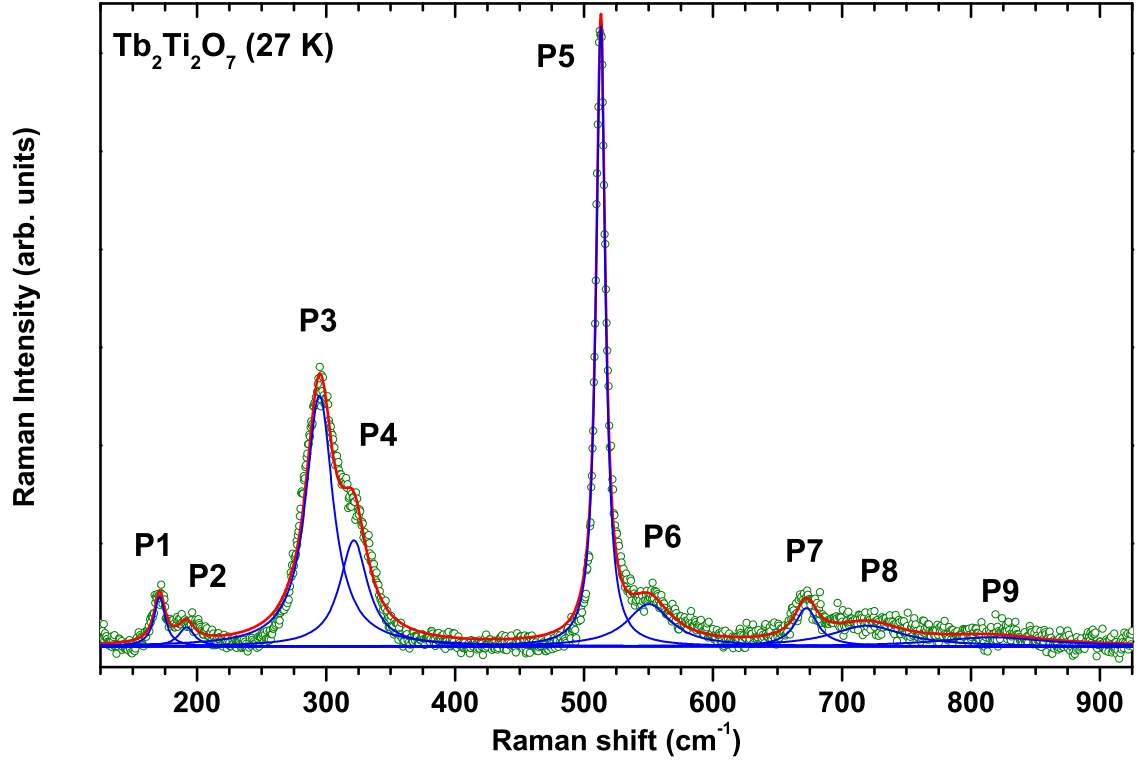} 
\caption{(Color online) Raman spectrum of Tb$_2$Ti$_2$O$_7$ at 27 K. Open circles represent the experimental data. Thin (blue) solid lines are the individual modes and thick (red) line is the total fit to the experimental data. Assignment of the modes P1 to P9 are done in the text (Table-I).} \label{Fig:1}
\end{center}
\end{figure}

Pyrochlores belong to the space group $Fd\bar{3}m (O^{h}_{7})$ with an $A_2B_2O_6O^{\prime}$ stoichiometry, where $A^{3+}$ occupies the 16d and $B^{4+}$ occupies the 16c Wyckoff positions and the oxygen atoms O and O$^{\prime}$ occupy the 48$f$ and 8$b$ sites, respectively. Factor group analysis for this family of structures gives six Raman active modes ($A_{1g}+E_g+4F_{2g}$) and seven infrared active modes ($7F_{1u}$). Raman spectra of Tb$_2$Ti$_2$O$_7$ have been recorded between 125 to 925 cm$^{-1}$ from room temperature down to 27 K. A strong Rayleigh contribution made the signal to noise ratio poor below 125 cm$^{-1}$. Fig. \ref{Fig:1} shows the Raman spectrum at 27 K, fitted with Lorentzians and labeled as P1 to P9. Following previous reports \cite{APL-86-181906, PRB-78-134420, PRB-78-214102, PRB-77-214310, JRS-39-537, CPL-413-248, JPCB-106-4663, JRS-14-63, JRS-32-41}, the modes can be assigned as follows: P3 (294 cm$^{-1}$, $F_{2g}$), P4 (325 cm$^{-1}$, $E_g$), P5 (513 cm$^{-1}$, $A_{1g}$) and P6 (550 cm$^{-1}$, $F_{2g}$). One $F_{2g}$ mode near 425 cm$^{-1}$ (observed in other pyrochlore titanates \cite{JRS-39-537,PRB-78-134420, PRB-78-214102}) could not be observed due to weak signal. The mode P1 (170 cm$^{-1}$) has been assigned to be the fourth $F_{2g}$ mode by refs. \cite{APL-86-181906, CPL-413-248, JPCB-106-4663, JRS-39-537, PRB-77-214310, PRB-78-134420, PRB-78-214102, JRS-14-63, JRS-32-41}. However, there has been a controversy on the assignment of the P1 mode \cite{DTO18, JINC-38-1407, AnnChim-9-43} and we, therefore, assign the mode P7 (672 cm$^{-1}$) as the fourth $F_{2g}$ mode. We support this assignment for the following reason: it is well established that the symmetry-allowed six Raman active modes ($A_{1g}+E_g+4F_{2g}$) in pyrochlore involve only the vibrations of oxygen atoms. This will imply that isotopic substitution by O$^{18}$ in pyrochlore should lower the phonon frequencies by $\sim$ 5 \%. This has, indeed, been seen in our recent experiments \cite{DTO18} on Dy$_2$Ti$_2$O$_7$ and Lu$_2$Ti$_2$O$_7$ for the modes P3 to P9 but not for P1 and P2. Another argument against P7 being a combination mode is the pressure dependence of the modes presented later (Fig. \ref{Fig:7}). Possible candidates for the combination are $\omega_{P7}\approx \omega_{P3}+\omega_{P4}$ and $\omega_{P7}\approx \omega_{P1}+\omega_{P5}$. The pressure derivative of frequency of the mode P7 ($\frac{d\omega_{P7}}{dP}$) does not agree with the sum of the pressure derivatives of the individual modes. Next, the question arises on the origin of the modes P1 and P2. Since these modes are also seen in Gd$_2$Ti$_2$O$_7$ and in non-magnetic Lu$_2$Ti$_2$O$_7$, their crystal field (CF) origin can be completely ruled out. We, therefore, attribute these low frequency modes to disorder induced Raman active modes. The high frequency modes (P8 and P9) are possibly second-order Raman modes \cite{JRS-39-537, PRB-78-134420, PRB-78-214102}.

\subsection{Temperature dependence of phonons}

\begin{figure}
\begin{center}
\leavevmode
\includegraphics[width=0.8\textwidth]{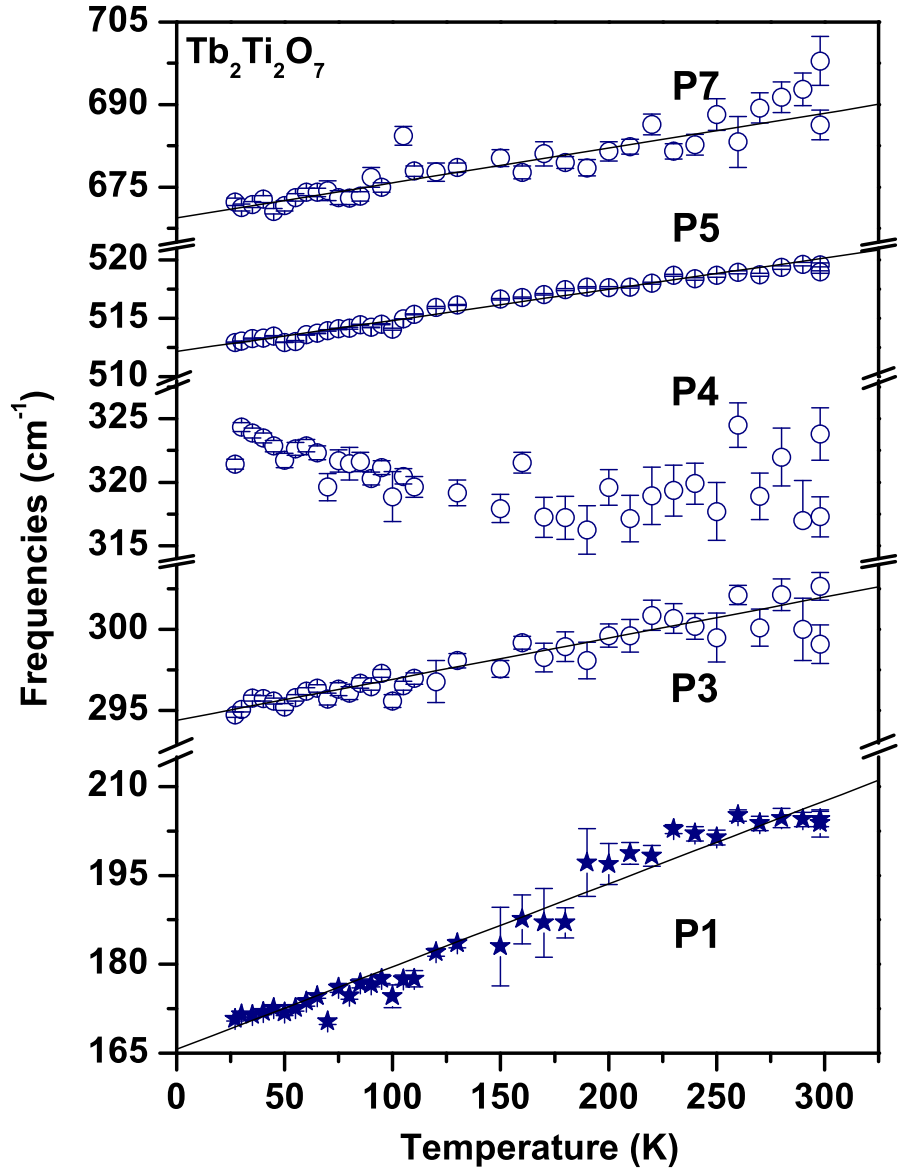}
\caption{(Color online) Temperature dependence of the modes P1, P3, P4, P5 and P7 of Tb$_2$Ti$_2$O$_7$. Solid lines are linear fit to the data. The slopes ($\frac{d\omega}{dT}$) of the corresponding modes are listed in the Table-I.} \label{Fig:2}
\end{center}
\end{figure}

We have recorded Raman spectra of Tb$_2$Ti$_2$O$_7$ from room temperature down to 27 K and followed the temperature dependence of the modes P1, P3, P4, P5 and P7. As shown in Fig. \ref{Fig:2}, the modes P1, P3, P5 and P7 soften with decreasing temperature. Since the Raman bands P2 and P6 are weak near room temperature, their temperature dependence is not shown. It needs to be mentioned that temperature-dependent anomalies of the modes P1, P5 and P7 have also been reported in other pyrochlore titanates \cite{PRB-77-214310, JRS-39-537, PRB-78-214102} and attributed to phonon-phonon anharmonic interactions. However, anomalous behavior of the $F_{2g}$(P3) mode near 300 cm$^{-1}$ has been reported only in the non-magnetic  Lu$_2$Ti$_2$O$_7$ pyrochlore \cite{PRB-78-214102}. We evidence a similar anomaly in P3 in Tb$_2$Ti$_2$O$_7$ with unusually broad linewidth. Recently, Maczka \textit{et al.} \cite{PRB-78-134420} have also reported this unusually broad linewidths in  Tb$_2$Ti$_2$O$_7$ which has been explained in terms of coupling between phonon and crystal field transition.

\begin{figure}
\begin{center}
\leavevmode
\includegraphics[width=0.7\textwidth]{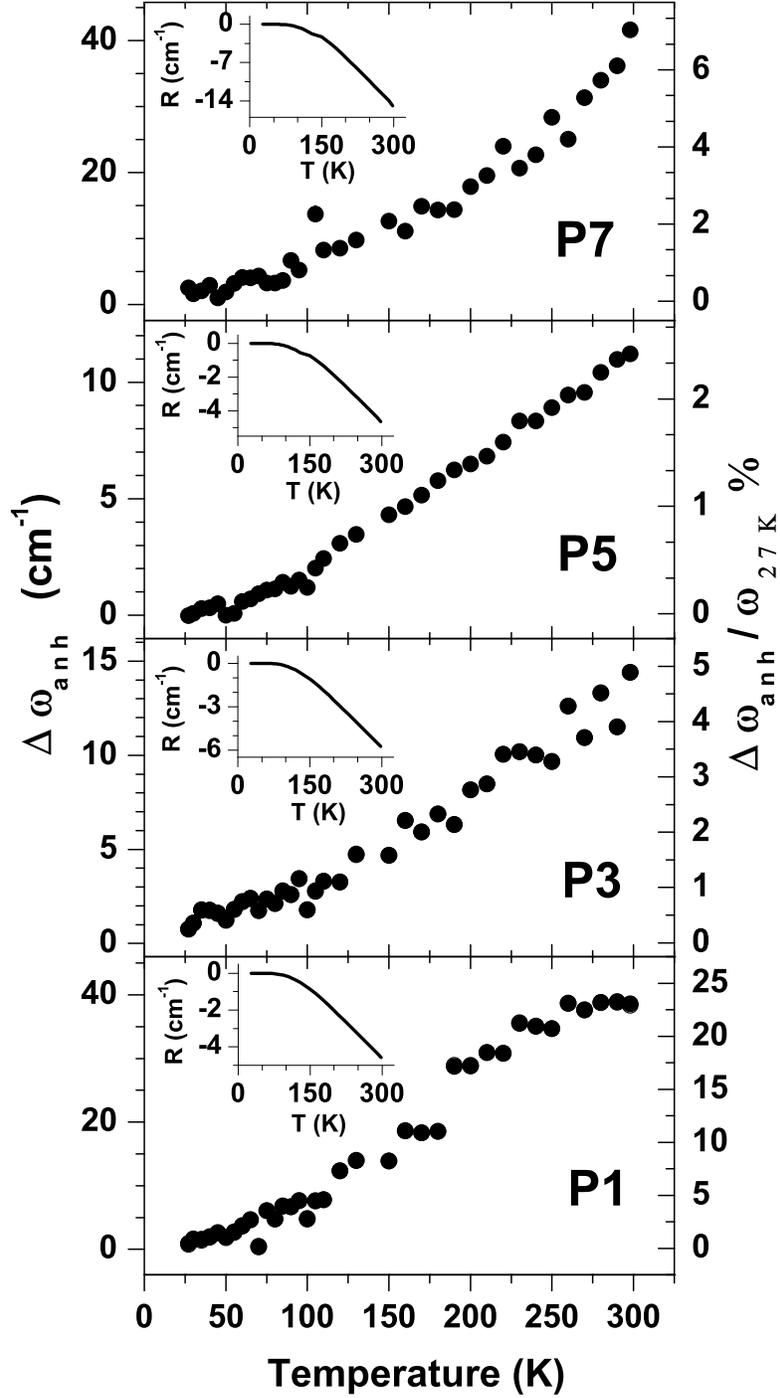}
\caption{The temperature-dependent anharmonic contribution to the phonons P1, P3, P5 and P7 estimated as described in text. The insets show the quasiharmonic changes (R=$(\Delta \omega_i)_{qh} (T) \approx (\omega_i)_{qh}(T) - \omega_i(27K)$) in the corresponding phonons upon changing temperature (T).} \label{Fig:3}
\end{center}
\end{figure}

Temperature dependence of a phonon mode ($i$) of frequency $\omega_i(T)$ can be expressed as \cite{PRB-28-1928},
\begin{eqnarray}
\omega_i(T)= && \omega_i(0)+{(\Delta \omega_i)}_{total}(T) \nonumber \\
\text{where,}\hspace {5 mm} {(\Delta \omega_i)}_{total}(T) = && {(\Delta \omega_i)}_{qh}(T)+{(\Delta \omega_i)}_{anh}(T) \nonumber \\
&& +{(\Delta \omega_i)}_{el-ph}(T)+{(\Delta \omega_i)}_{sp-ph}(T)
\end{eqnarray}
The term $\omega_i(0)$ corresponds to the phonon frequency at absolute zero. In eqn. 1 above, the first term on the right hand side corresponds to quasiharmonic contribution to the frequency change. The second term corresponds to the intrinsic anharmonic contribution to phonon frequency that comes from the real part of the self-energy of the phonon decaying into two phonons (cubic anharmonicity) or three phonons (quartic anharmonicity). The third term ${(\Delta \omega_i)}_{el-ph}$ is the renormalisation of the phonon energy due to coupling of phonons with charge carriers in the system which is absent in insulating pyrochlore titanates. The last term, ${(\Delta \omega_i)}_{sp-ph}$, is the change in phonon frequency due to spin-phonon coupling arising from modulation of the spin exchange integral by the lattice vibration. Recently, we have shown \cite{PRB-78-214102} that the magnitude of phonon anomalies is comparable in both magnetic and non-magnetic pyrochlore titanates, thus ruling out any contribution from spin-phonon coupling. Therefore, the change in phonon frequency is solely due to quasiharmonic and intrinsic anharmonic effects whose temperature variations, as estimated below for the modes P1, P3, P5 and P7, are shown in Fig. \ref{Fig:3}.

\begin{figure}
\begin{center}
\leavevmode
\includegraphics[width=0.8\textwidth]{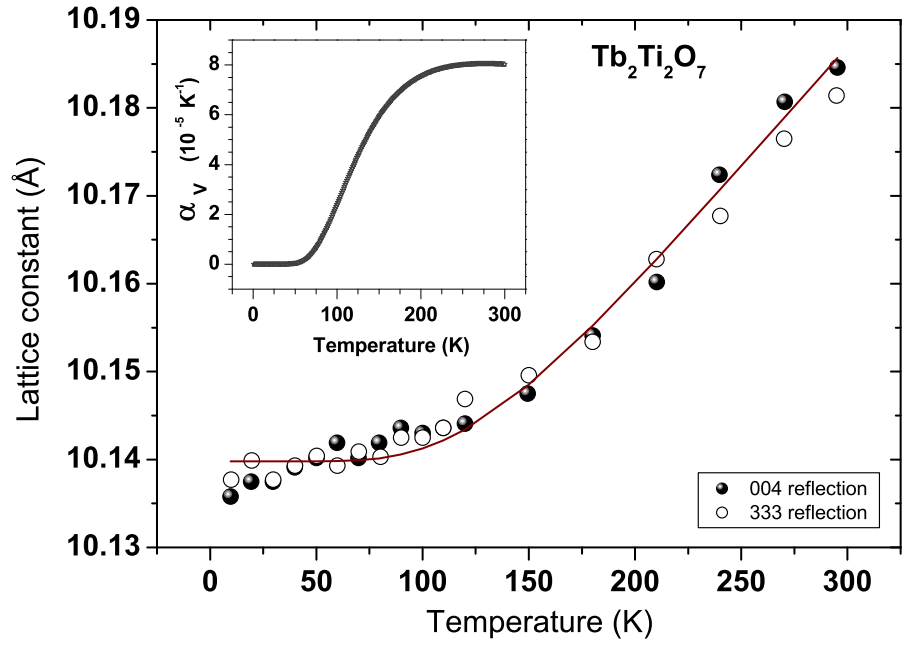}
\caption{(Color online) Variation of the lattice parameter of Tb$_2$Ti$_2$O$_7$ with temperature. Open and closed circles are  values obtained from (333) and (004) diffraction peaks of our x-ray data. Solid line is the fit to the data as discussed in text. Inset shows the coefficient of thermal expansion ($\alpha_V$) derived from the lattice parameter.} \label{Fig:4}
\end{center}
\end{figure}

The change in phonon frequency due to quasiharmonic effects ($(\Delta \omega_i)_{qh} (T)$) comes from the change in the unit cell volume. This change can be expressed as \cite{Born-n-Huang}, 
\begin{equation}
{(\omega_i)}_{qh}(T) - \omega_i(0) = (\Delta \omega_i)_{qh} (T) = \omega_i(0) \hspace{1.5 mm} exp\left({\displaystyle\int^{T}_{0} \gamma_i(T^{\prime}) \alpha_v(T^{\prime})\, dT^{\prime}}\right)
\end{equation}
where $\omega_i(0)$ is the frequency of the $i^{th}$ phonon mode at 0 K, $\gamma_i(T^{\prime})$ is the temperature-dependent Gr\"{u}neisen parameter of that phonon and $\alpha_v(T^{\prime})$ is the temperature-dependent coefficient of the volume expansion. Since our lowest temperature is 27 K, the quasiharmonic change can be approximated as, 
\begin{equation}
(\Delta \omega_i)_{qh} (T) \approx \omega_i(27K) \hspace{1.5 mm} \gamma_i \hspace{1.5 mm} exp \left({\int^{T}_{27K} \alpha_v(T^{\prime})\, dT^{\prime}}\right) 
\end{equation}
assuming the Gr\"{u}neisen parameter to be temperature independent. To measure the $\alpha_v(T)$, we have recorded x-ray diffraction patterns of Tb$_2$Ti$_2$O$_7$ from room temperature to 10 K. We present the temperature-dependent lattice parameter in Fig. \ref{Fig:4}. Our data agree with the recent data by Ruff \textit{et al.} \cite{PRL-99-237202}. The solid line in Fig. \ref{Fig:4} is a fit to our data by the relation $a(T)=a_0[1+\frac{be^{c/T}} {T(e^{c/T}-1)^2 }]$, where $a_0$=10.14 $\AA$ is the lattice constant at 0 K and b=9.45 K and c=648.5 K are fitting parameters \cite{KittleBook}. In a recent study by Ruff et al. \cite{PRL-99-237202}, it was shown that the lattice undergoes an anomalous expansion along with broadening of allowed Bragg peaks as temperature is reduced below $\sim$ 10 K. This was attributed to structural fluctuation from cubic-to-tetragonal lattice that consequently coincides with the development of correlated spin-liquid ground state in Tb$_2$Ti$_2$O$_7$. Our data are up to 10 K and hence, we could not observe this feature at low temperatures. We have derived the temperature-dependent coefficient of thermal expansion ($\alpha_v = \frac{3}{a_0} \frac{da}{dT}$) from the temperature-dependent lattice parameter which is shown in the inset of Fig. \ref{Fig:4}. The $\alpha_v$ at 300 K for Tb$_2$Ti$_2$O$_7$ is ($\sim 8\times10^{-5} K^{-1}$), slightly higher than those of Dy$_2$Ti$_2$O$_7$ ($\sim 6\times10^{-5}K^{-1}$) and Lu$_2$Ti$_2$O$_7$ ($\sim 4\times10^{-5}K^{-1}$), estimated from the temperature-dependent x-ray diffraction results reported in ref. \cite{PRB-78-214102}. We note that $\alpha_v(300K)$ for Tb$_2$Ti$_2$O$_7$ is about 10 times higher than that of Si \cite{JAP-56-314} and nearly 7 times higher than that of Gd$_2$Zr$_2$O$_7$ \cite{PML-84-127}, implying that the anharmonic interactions in Tb$_2$Ti$_2$O$_7$ are strong. The mode Gr\"{u}neisen parameter for $i^{th}$ phonon mode is $\gamma_i=\frac{B}{\omega_i} \frac{d\omega_i}{dP}$, where $B$ is the bulk modulus, $\frac{d\omega_i}{dP}$ is the frequency change with pressure $P$. Taking B=154 GPa, obtained from our high pressure x-ray diffraction data discussed later, we find the values of the Gr\"{u}neisen parameter for the various modes as listed in Table-I. The change in phonon frequency due to quasiharmonic effect, $(\Delta \omega_i)_{qh}(T)$, has been estimated for the modes P1, P3, P5 and P7, and is shown in the insets of Fig. \ref{Fig:3}. The anharmonic contribution, $(\Delta \omega_i)_{anh}(T)=(\Delta \omega_i)_{total}(T)-(\Delta \omega_i)_{qh}(T)$, for the modes P1, P3, P5 and P7 are shown in Fig. \ref{Fig:3}. We note that the temperature-dependent $(\Delta \omega_i)_{anh}(T)$ for these four modes is anomalous. Further, upon changing the temperature from 27 K to 300 K, we find that for the mode P1, the percentage change in frequency due to anharmonic interactions, $(\Delta \omega_i)_{anh}(T)/\omega_i(27K)$, is exceptionally high. It is customary to fit the $(\Delta \omega_i)_{anh}(T)$ data by the expression \cite{PRB-28-1928},
\begin{equation}
(\Delta \omega_i)_{anh}(T) = C \left(1+\frac{2}{e^{\frac{\bar{h}\omega_i(0)}{2k_BT}}-1}\right)
\end{equation}
where the $i^{th}$ phonon decays into two phonons of equal energy ($\omega_i \rightarrow \frac{\omega_i}{2} + \frac{\omega_i}{2}$). The parameter ``$C$'' can be positive (for normal behavior of phonon) or negative (anomalous phonon) \cite{PRB-78-214102, SSC-117-201}. We have seen that eqn. 4 does not fit to our data of $(\Delta \omega_i)_{anh}(T)$ (fitting not shown in Fig. \ref{Fig:3}). This may be because, in the expression for $(\Delta \omega_i)_{anh}(T)$ (eqn. 4), all the decay channels for the phonons are not taken into account. Therefore, a full calculation for the anharmonic interactions considering all the possible decay channels is required to understand the $(\Delta \omega_i)_{anh}(T)$ data, shown in Fig. \ref{Fig:3}.

\begin{figure}
\begin{center}
\leavevmode
\includegraphics[width=0.75\textwidth]{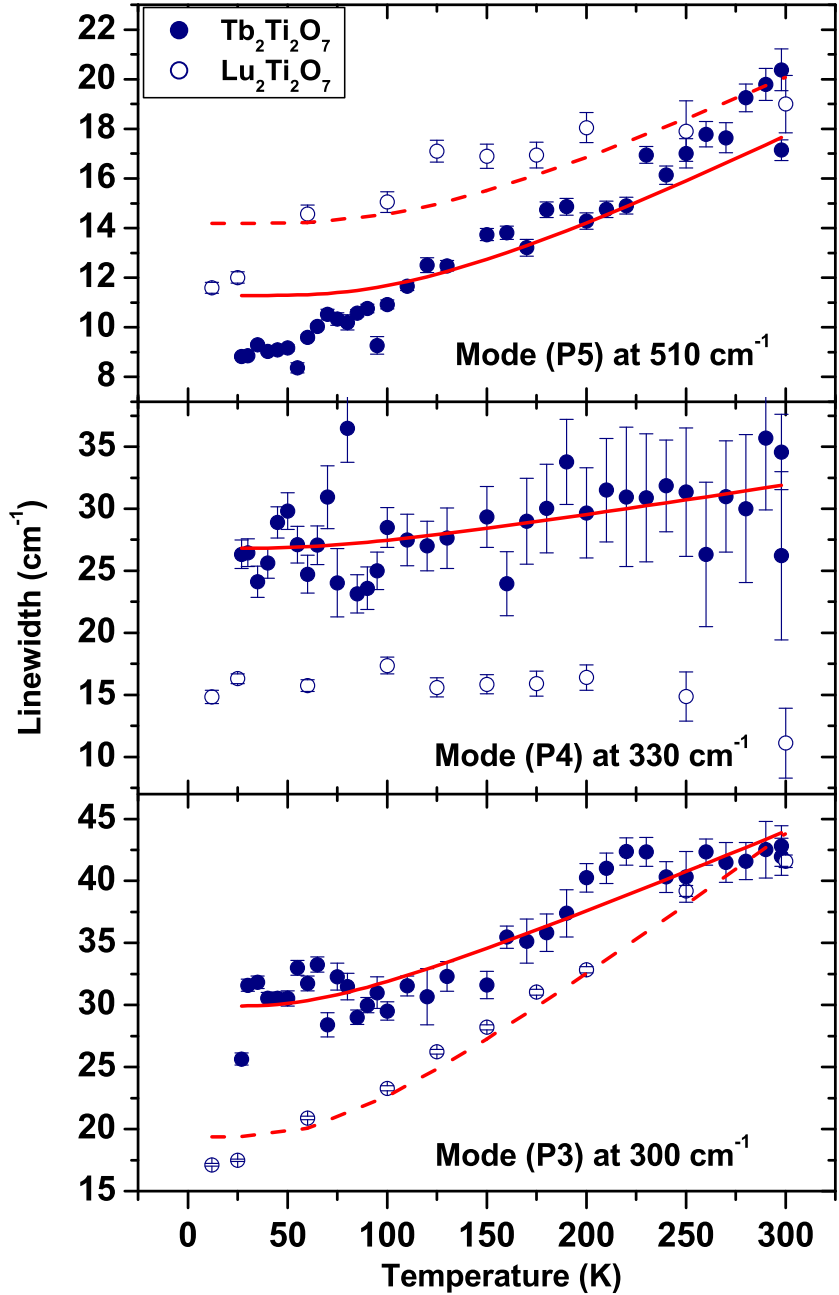}
\caption{(Color online) Closed circles are the experimental values of temperature-dependent linewidths of the modes P3, P4 and P5 of Tb$_2$Ti$_2$O$_7$. They have been compared with those of the same modes in non-magnetic Lu$_2$Ti$_2$O$_7$ (open circles) as discussed in the text. Solid and dashed lines are fits to the data of Tb$_2$Ti$_2$O$_7$ and Lu$_2$Ti$_2$O$_7$, respectively, as explained in the text.} \label{Fig:5}
\end{center}
\end{figure}

Considering only the cubic phonon-phonon anharmonic interactions where a phonon decays into two phonons of equal energy, the temperature-dependent broadening of the linewidth can be expressed as \cite{PRB-28-1928}:
\begin{equation}
\Gamma_i(T) = \Gamma_i(0) + A \left(1+\frac{2}{e^{\frac{\bar{h}\omega_i(0)}{2k_BT}}-1}\right)
\end{equation}
where $\omega_i(0)$ is the zero temperature frequency and $\Gamma_i(0)$ is the linewidth arising from disorder. Fig. \ref{Fig:5} shows the temperature dependence of linewidths of the Raman modes P3, P4 and P5. It can be seen that the linewidth of P3 and P4 modes are almost double of the linewidth of the P5 mode, as reported by Maczka \textit{et al.} \cite{PRB-78-134420}. These authors have attributed this to the strong coupling of the $F_{2g}$(P3) and $E_g$(P4) phonons with the crystal field transitions of Tb$^{3+}$ which is absent for the $A_{1g}$ (P5) mode due to symmetry consideration. To strengthen this argument, we compare (Fig. \ref{Fig:5}) these results with the linewidths of the corresponding phonons in non-magnetic Lu$_2$Ti$_2$O$_7$ (Lu$^{3+}$: J=0) \cite{PRB-78-214102} and, indeed, the linewidths of P3 and P4 modes in Tb$_2$Ti$_2$O$_7$ are much broader than those in Lu$_2$Ti$_2$O$_7$. The change in linewidth of the $F_{2g}$(P3) mode in Tb$_2$Ti$_2$O$_7$ from room temperature down to 27 K is nearly half the change in linewidth of the same mode in Lu$_2$Ti$_2$O$_7$ and, therefore, the parameter ``$A$'' for $F_{2g}$(P3) mode in Tb$_2$Ti$_2$O$_7$ is 7.2 cm$^{-1}$, which is nearly half of that in Lu$_2$Ti$_2$O$_7$ ($A$=13.4 cm$^{-1}$). However, the linewidth of the $A_{1g}$ mode for both titanates is comparable and, therefore, the fitting parameter ``$A$'' for this mode in Tb$_2$Ti$_2$O$_7$ and Lu$_2$Ti$_2$O$_7$ are nearly the same, i.e., 7.8 cm$^{-1}$ and 7.3 cm$^{-1}$, respectively. All these results, therefore, corroborate the suggestion of Maczka \textit{et al.} \cite{PRB-78-134420} thus emphasizing a strong coupling between the phonon and crystal field modes.

\subsection{Effect of pressure on Tb$_2$Ti$_2$O$_7$}

\subsubsection{Raman study}

\begin{figure}
\begin{center}
\leavevmode

\includegraphics[width=0.75\textwidth]{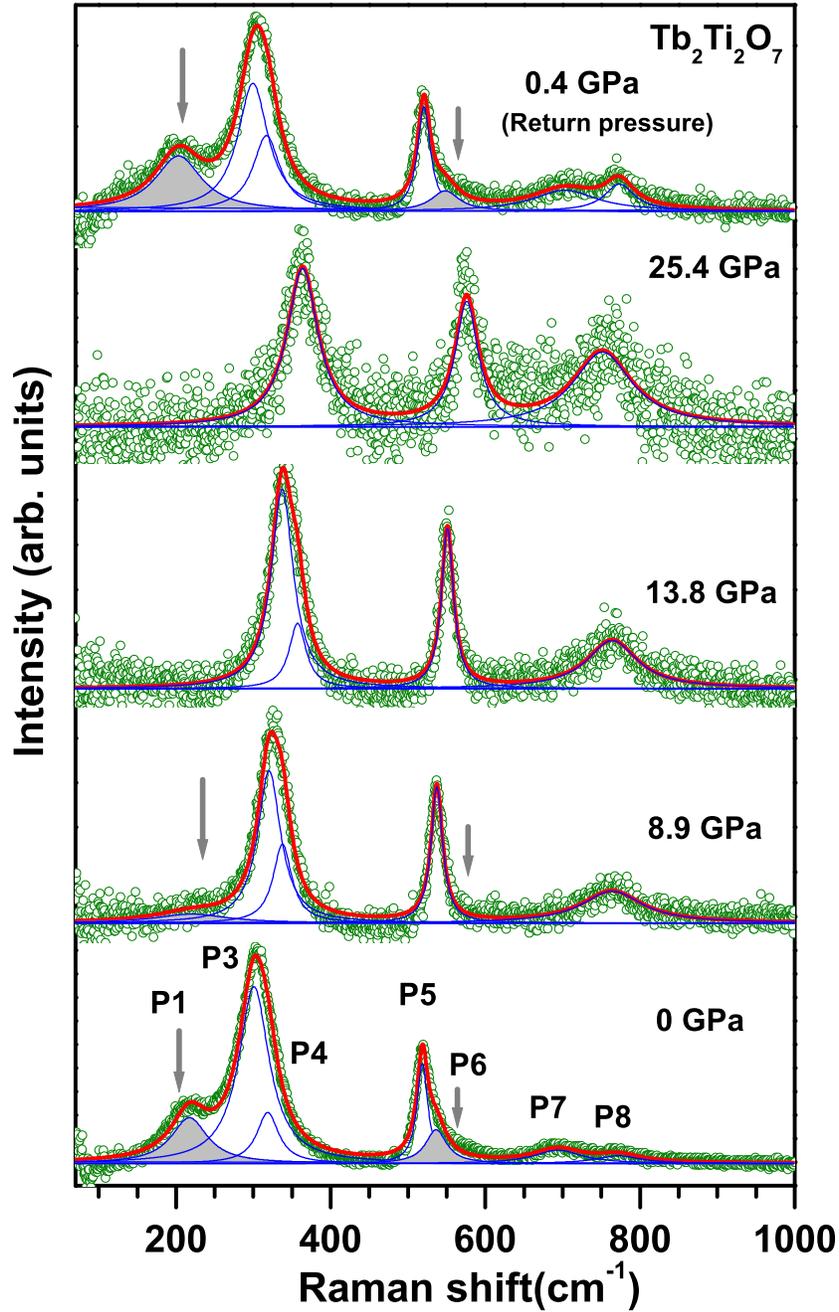}
\caption{(Color online) Raman spectra of Tb$_2$Ti$_2$O$_7$ at ambient and a few high pressures. Open circles are the experimental data. Thin (blue) solid lines are the fits for individual modes and the thick (red) solid line is total fit to the data. The modes P1 and P6 disappears above $\sim$ 9 GPa and reappear on decompression as indicated by the arrows.} \label{Fig:6}
\end{center}
\end{figure}

\begin{figure}
\begin{center}
\leavevmode
\includegraphics[width=0.75\textwidth]{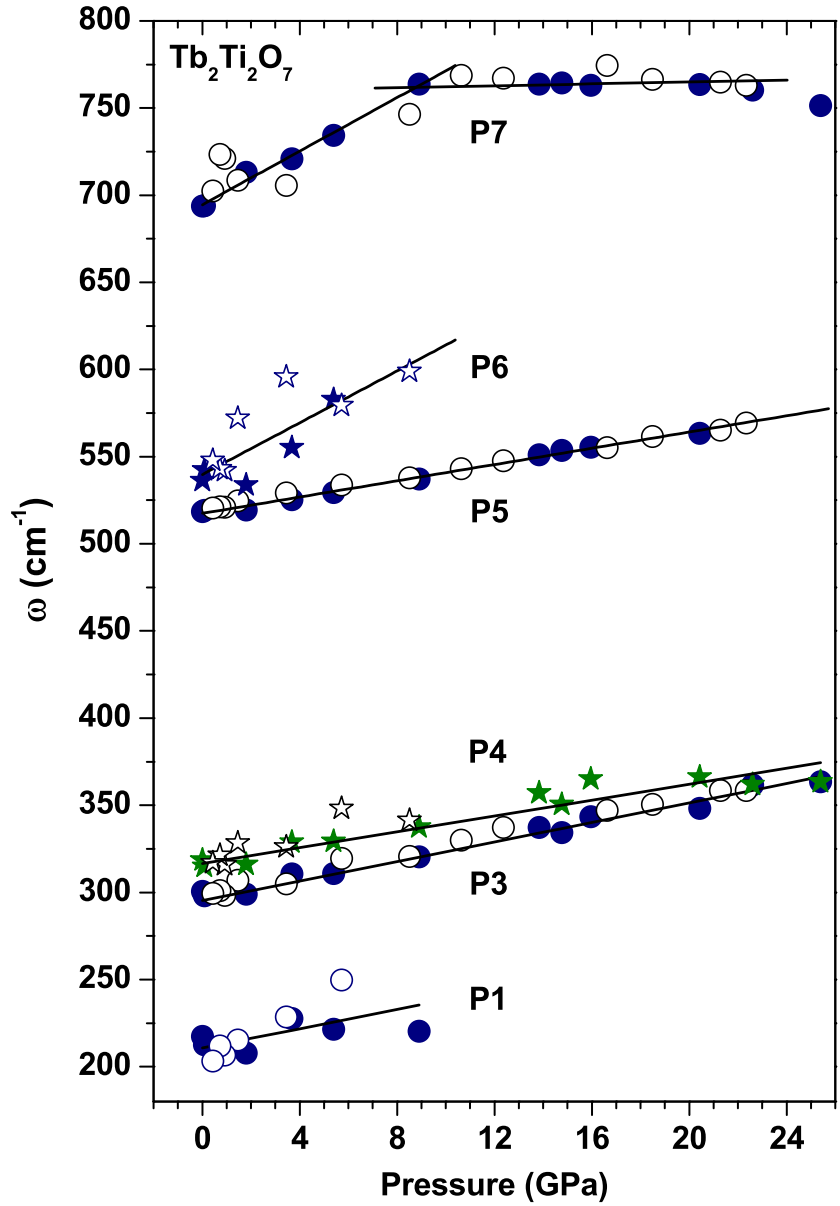}
\caption{(Color online) Pressure-dependent frequencies of the modes P1, P3, P4, P5, P6 and P7. Filled and open symbols correspond to the compression and decompression data. Solid lines are linear fits to the frequencies. Modes P1 and P6 disappear above 9 GPa and the mode P7 undergoes a dramatic change in slope ($\frac{d\omega}{dP}$) at the same pressure thus suggesting a subtle structural transition. The slopes are listed in Table-I.} 
\label{Fig:7}
\end{center}
\end{figure}

\begin{figure}
\begin{center}
\leavevmode
\includegraphics[width=0.7\textwidth]{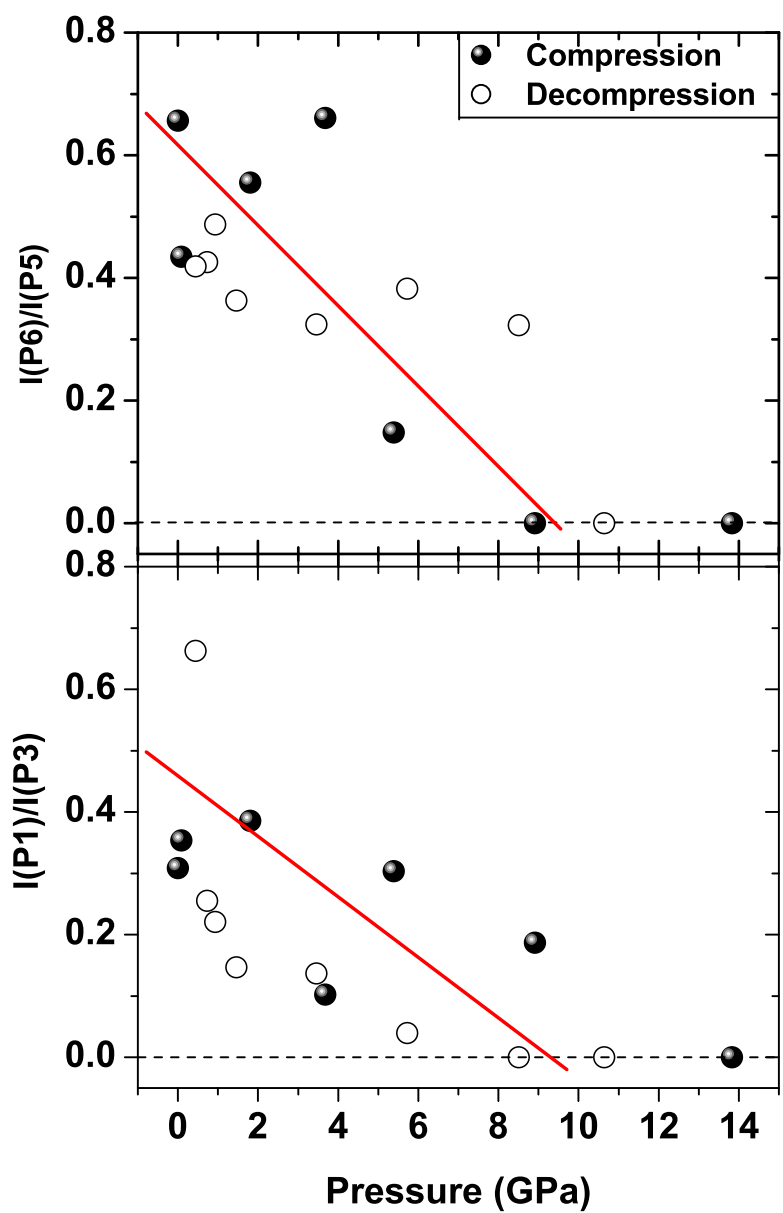}
\caption{(Color online) Intensity ratio of the modes P1 to P3 (bottom panel) and P6 to P5 (top panel) vs pressure thus suggesting that the intensities of the modes P1 and P6 gradually decrease with increasing pressure and the modes disappear above $\sim$ 9 GPa. Closed and open circles correspond to the compression and decompression data, respectively, and solid lines are guide to eye.} \label{Fig:8}
\end{center}
\end{figure}

Fig. \ref{Fig:6} shows room temperature Raman spectra at ambient and a few high pressures, the maximum pressure being $\sim$ 25 GPa. We could not resolve P2 and P9 at room temperature inside the high pressure cell due to the reasons described above. The phonon frequencies increase with increasing pressure, as shown in Figs. \ref{Fig:6} and \ref{Fig:7}. Interestingly, we find that upon increasing the pressure, the intensity of the P1 mode diminishes and is no longer resolvable above $\sim$ 9 GPa. On decompressing the sample from $\sim$ 25 GPa, the mode recovers, as shown in the top panel of Fig. \ref{Fig:6}. Similarly the mode P6 ($F_{2g}$) also vanishes above $\sim$ 9 GPa and reappears on decompression. The intensity ratios of the modes P1 to P3 and P6 to P5, as shown in Fig. \ref{Fig:8}, gradually decrease with increasing pressure and become zero near 9 GPa. As shown in Fig. \ref{Fig:7}, the maximum change in phonon frequency is seen in mode P7 ($F_{2g}$), which shows a dramatic change in the rate of change of frequency with pressure at a pressure of $\sim$ 9 GPa. In sharp contrast, the other modes P3, P4 and P5 do not show any change in slope till the maximum pressure applied. The changes seen in the modes P1, P6 and P7 near 9 GPa are indicative of a structural transition of the Tb$_2$Ti$_2$O$_7$ lattice. In order to ascertain the structural transition we have performed high pressure x-ray diffraction measurements and the results are discussed below.

\subsubsection{X-ray diffraction}

\begin{figure}
\begin{center}
\leavevmode
\includegraphics[width=0.8\textwidth]{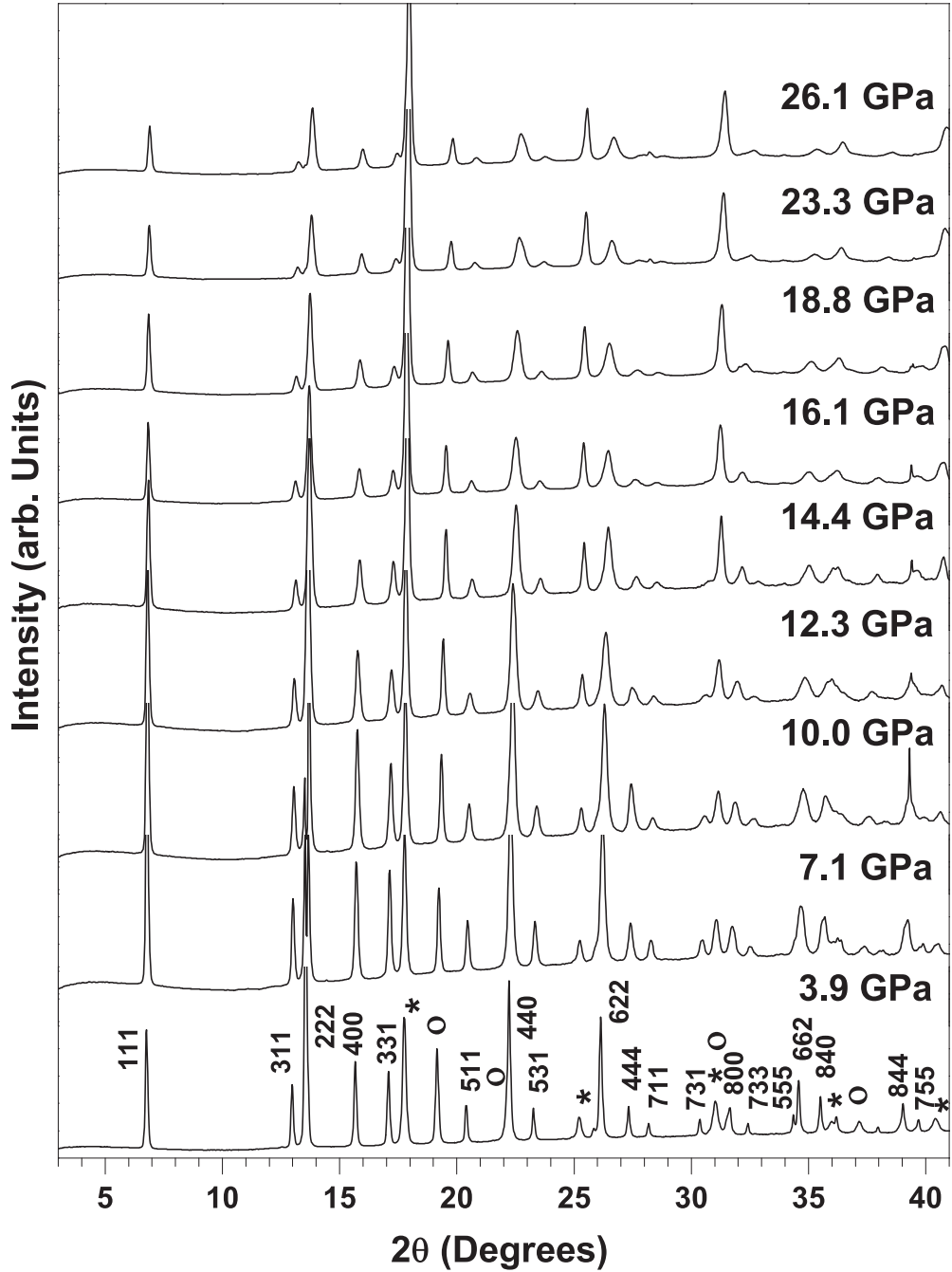}
\caption{X-ray diffraction patterns of pyrochlore Tb$_2$Ti$_2$O$_7$ at a few high pressures. (h k l) values are marked for each of the diffraction peaks. Stars (*) and open circles (O) represent tungsten gasket and Cu (pressure marker) peaks, respectively.} \label{Fig:9}
\end{center}
\end{figure}

\begin{figure}
\begin{center}
\leavevmode
\includegraphics[width=0.85\textwidth]{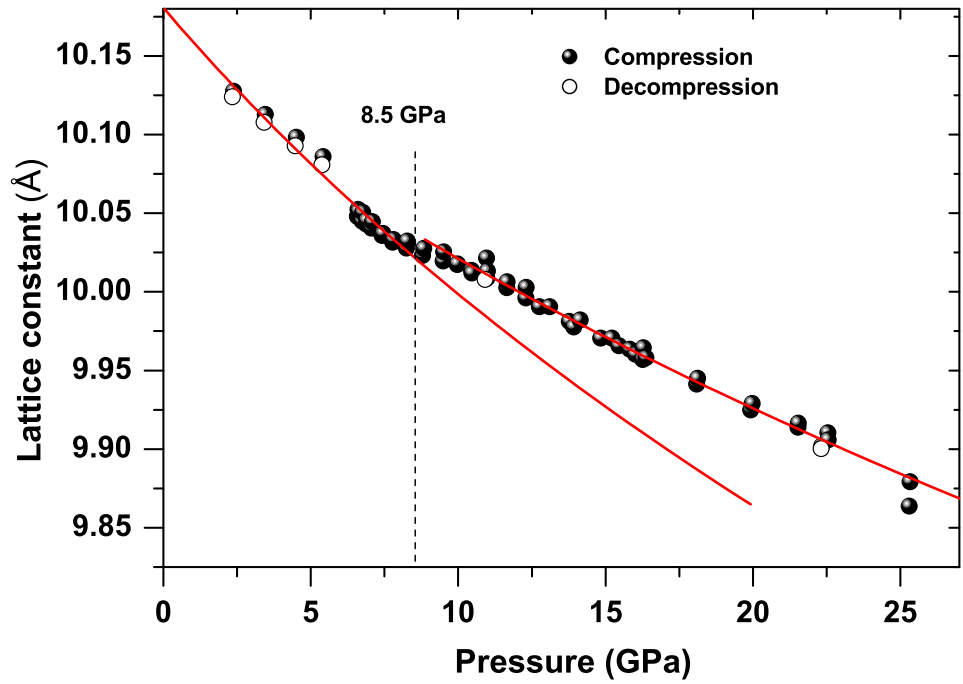}
\caption{Pressure dependence of lattice parameter. Open and closed symbols represent the compression and decompression data. Solid lines are fit to the data as discussed in text.} \label{Fig:10}
\end{center}
\end{figure}

Fig. \ref{Fig:9} shows the x-ray diffraction patterns of Tb$_2$Ti$_2$O$_7$ at a few high pressures. The (hkl) values are marked on the corresponding diffraction peaks. As we increase the pressure, we find that the diffraction peaks shift to higher angles but no signature of new peak or peak splitting could be observed. However, the change in lattice parameter with pressure, shown in Fig. \ref{Fig:10}, shows a change in slope near 9 GPa implying a structural deformation, thus corroborating the transition observed in the Raman data. The transition possibly involves just a local rearrangement of the atoms retaining the cubic symmetry of the crystal. Fitting the pressure-dependent volume to the third order Birch-Murnaghan equation of state \cite{JGeophysRes-83-1257}, we find that B = 154 GPa and B$^{\prime}$=6.6 when the applied pressure is below 9 GPa. But, when the applied pressure is above this transition pressure, these values change to B = 250 GPa and B$^{\prime}$ = 7.1 thus implying an increment of the bulk modulus by $\sim$ 62\% after the transition. A similar transition had also been observed \cite{PRB-74-064109} in Gd$_2$Ti$_2$O$_7$ at $\sim$ 9 GPa and was attributed to the TiO$_6$ octahedral rearrangement. It needs to be mentioned here that the pressure transmitting medium (methanol-ethanol mixture, used in our Raman experiments) remains hydrostatic up to 10 GPa which is close to the transition pressure, thus implying that the possibility of a contribution from non-hydrostaticity of the medium cannot be completely ruled out. However, experiments in a non-hydrostatic medium (water) has as well revealed the transition at $\sim$ 9 GPa in Gd$_2$Ti$_2$O$_7$ \cite{PRB-74-064109}. We, therefore, believe that the transition near 9 GPa is an intrinsic property of Tb$_2$Ti$_2$O$_7$ and also that performing this experiment with helium as the pressure transmitting medium, will further strengthen our suggestion of a possible transition at $\sim$ 9 GPa.

\section{Summary and Discussion}

We have performed temperature and pressure-dependent Raman and x-ray diffraction studies on pyrochlore Tb$_2$Ti$_2$O$_7$ and the main results can be summarised as follows: (1) The phonon frequencies show anomalous temperature dependence, (2) the linewidths of the $F_{2g}$ and $E_g$ modes near 300 cm$^{-1}$ are unusually broad in comparison to those of non-magnetic Lu$_2$Ti$_2$O$_7$ phonons, thus corroborating the suggestion \cite{PRB-78-134420} of a possible coupling between phonons and crystal field transitions, (3) intensities of two phonon modes (P1 and P6) decrease to zero as the applied pressure approaches 9 GPa. Another Raman band P7 near 672 cm$^{-1}$ ($F_{2g}$) shows a large change in slope ($\frac{d\omega}{dP}$) at $\sim$ 9 GPa, thus indicating a possible transition, (4) x-ray diffraction study as a function of pressure reveals an increase in bulk modulus by $\sim$ 62\% when the applied pressure is above 9 GPa thus corroborating the transition suggested by Raman data. The phonons in Tb$_2$Ti$_2$O$_7$ show anomalous temperature dependence which has been attributed to the phonon-phonon anharmonic interactions \cite{PRB-78-214102}. Using the required parameters ($\gamma$, B and $\alpha_v$), derived from our high pressure and temperature-dependent Raman and x-ray experiments, we have estimated the contributions of quasiharmonic and anharmonic effects (Fig. \ref{Fig:3}) to the phonon frequencies. We note that the anharmonicity of the mode P1 (mode near 200 cm$^{-1}$) is unusually high as compared to other modes. P1 is a phonon mode which do not involve oxygen but includes the vibrations of Ti$^{4+}$ ions \cite{DTO18}. This can be qualitatively understood by examining how Ti$^{4+}$ and Tb$^{3+}$ ions are coordinated. There are tetrahedra in the unit cell which are occupied by Ti$^{4+}$ ions at the vertices with a vacant 8$a$-site inside. The later will tend to make the vibrational amplitudes of Ti$^{4+}$ ions larger and thus contributing to the high anharmonic nature of the P1 mode. The high anharmonic behavior of the Raman modes involving 48$f$-oxygen ions arises due to the fact that the O$_{48f}$ anions are off centered towards the 8$a$-vacant site from their ideal position $\frac{3}{8}a$ to $(\frac{3}{8}-x)a$ inside the tetrahedra \cite{PSSC-15-55} whose two vertices are occupied by Ti$^{4+}$ and other two by Tb$^{3+}$. Here $a$ is the lattice parameter and $x$ is the O$_{48f}$ positional parameter. This anharmonicity is reflected in the high root mean squared displacement ($\sqrt{\langle u^2 \rangle}$) of O$_{48f}$ atoms : $\frac{\sqrt{\langle u^2 \rangle}}{d_{Tb-O}} \approx 3\%$ \cite{PRB-69-024416}, where $d_{Tb-O}$ is the Tb-O bond length.

Pressure-dependent Raman data show that two Raman modes, P1 and P6, cannot be seen above $P_c \sim$ 9 GPa and the P7 ($F_{2g}$) Raman band shows a significant change in the slope ($\frac{d\omega}{dP}$) at $P_c$. These results suggest a subtle stuctural deformation which gets corroborated by a change in bulk modulus seen in pressure-dependent x-ray experiments. However, the pressure-dependent x-ray data do not reveal any new diffraction peak or splitting of line. This implies that the structural deformation near 9 GPa, as inferred from the Raman study, is a local distortion of the lattice. It may be possible that as pressure increases, due to the vacancies at the $8a$-sites, the Ti$^{4+}$-ions adjust their local coordinates with a concomitant relocation of other atoms in the lattice. At this instant, we would like to recall the results of a neutron scattering experiment on Tb$_2$Ti$_2$O$_7$ by Mirebeau \textit{et al.} with a simultaneous change in pressure and temperature \cite{Nature-420-54}. It was seen that at 1.5 K, antiferromagnetic correlations develop in Tb$_2$Ti$_2$O$_7$ at a pressure of 8.6 GPa. This was attributed to the delicate balance among the exchange coupling, crystal field and dipolar interactions that gets destroyed under high pressure. Our high pressure Raman and x-ray experiments on Tb$_2$Ti$_2$O$_7$ suggest a local rearrangement of the atoms near 9 GPa retaining the cubic symmetry which, we believe, may contribute to the antiferromagnetic correlations observed in neutron scattering experiments \cite{Nature-420-54}. The possibility of a structural transition in Tb$_2$Ti$_2$O$_7$ at low temperatures has recently been relooked. As discussed in section III(B), Ruff \text{et al.} \cite{PRL-99-237202} suggested an onset of cubic-to-tetragonal structual fluctuations below 20 K. A simultaneous presence of a CF mode at $\sim$ 13 cm$^{-1}$ in Raman and infrared spectrosopic measurements led Lummen \textit{et al.} \cite{PRB-77-214310} to propose a broken inversion symmetry in Tb$_2$Ti$_2$O$_7$ at low temperatures. The authors suggested the presence of a second Tb$^{3+}$ site with different site symmetry at low temperatures. Followed by this, Curnoe \cite{PRB-78-094418} has proposed that a structural transition can occur at low temperatures with an $A_{2u}$ lattice distortion resulting in a change of the point group symmetry, leaving the cubic lattice unchanged. Our Raman spectroscopic observations of a transition near 9 GPa may be related to the above discussion and can contribute to the increase in magnetic correlation observed by Mirebeau \textit{et al.} \cite{Nature-420-54}. It will be relevant to do high pressure Raman experiments at helium temperatures to strengthen our suggestion.

\section{Conclusion}

To conclude, our Raman spectroscopic and x-ray diffraction experiments on single crystals of pyorhclore Tb$_2$Ti$_2$O$_7$, with temperature, reveal highly anomalous temperature-dependent phonons attributed to strong phonon-phonon anharmonic interactions. Our pressure-dependent Raman and x-ray diffraction experiments suggest a local deformation of the pyrochlore lattice near 9 GPa. We believe that our experimental results play an important role in enriching the understanding of pyrochlore titanates, especially the spin-liquid Tb$_2$Ti$_2$O$_7$, thus motivating further experimental and theoretical studies on these exotic systems.

\begin{acknowledgments}
We thank the Indo-French Centre for Promotion of Advanced Research (IFCPAR), Centre Franco-Indien pour la Promotion de la Recherche Avanc\'{e}e (CEFIPRA) for financial support under Project No. 3108-1.  AKS thanks the Department of Science and Technology (DST), India, for partial financial support.
\end{acknowledgments}


\newpage

\begin{table}
\caption{A list of the phonon frequencies in cm$^{-1}$ at 27 K with the corresponding pressure and temperatures dependences.}
\begin{ruledtabular}
\begin{tabular}{lcccc}
Normal modes &$\omega$(cm$^{-1}$) &$\frac{d\omega}{dT}$(cm$^{-1}$/K) &$\frac{d\omega}{dP}$(cm$^{-1}$/GPa) \footnotemark[2] &Gr\"{u}neisen Parameter$(\gamma)$ \footnotemark[2] \\
\hline 
 P1 \footnotemark[1] (Phonon)      &170.8          &0.14             &2.8                                     &1.97  \\
 P2 \footnotemark[1] (Phonon)      &191.7          &0.17             &                                        &      \\
 P3 (Phonon, $F_{2g}$)             &294.8          &0.03             &2.8                                     &1.43  \\
 P4 (Phonon, $E_{g}$)              &321.4          &                 &2.3                                     &1.10  \\
 P5 (Phonon, $A_{1g}$)             &512.9          &0.03             &2.3                                     &0.69  \\
 P6 (Phonon, $F_{2g}$)             &550.3          &-0.004           &7.4                                     &2.13  \\
 P7 (Phonon, $F_{2g}$)             &672.3          &0.06             &7.7 (below 9 GPa)                       &1.70  \\
                                   &               &                 & and 0.3 (above 9 GPa)                  &      \\
 P8 \footnotemark[1] (Phonon)      &719.5          &                 &                                        &      \\
 P9 \footnotemark[1] (Phonon)      &816.5          &                 &                                        &      \\
\end{tabular}
\end{ruledtabular}
\footnotetext[1]{Origin of the mode has been discussed in the text.}
\footnotetext[2]{The values are at room temperature.}
\end{table}
\newpage


\end{document}